\documentclass[aps,prd,preprintnumbers,showpacs,twocolumn]{revtex4}
\usepackage{amsmath,amssymb}
\usepackage{graphicx}

\newcommand{\lp}{\ell_{\mathrm P}}
\newcommand{\be}{\begin{equation}}
\newcommand{\ee}{\end{equation}}
\newcommand{\bq}{\begin{eqnarray}}
\newcommand{\eq}{\end{eqnarray}}

\newcommand{\rcl}{\rho_{\mathrm{cl}}}
\newcommand{\rqm}{\rho_{\mathrm{q}}}
\newcommand{\rsc}{\rho_{\mathrm{sc}}}
\newcommand{\wcl}{w_{\mathrm{ cl}}}

\newcommand{\wsc}{w_{\mathrm{sc}}}
\newcommand{\wqm}{w_{\mathrm{q}}}

\newcommand{\lra}{\longrightarrow}
\begin{document}

\preprint{IGPG-05/2-2}

\title{Effective State Metamorphosis in Semi-Classical Loop Quantum Cosmology }
\author{Parampreet Singh\footnote{e-mail address: {\tt singh@gravity.psu.edu}}}
\affiliation{Institute for Gravitational Physics and Geometry, Pennsylvania State University,
University Park, PA 16802, USA}

\begin{abstract}
Modification to the behavior of geometrical density at short
scales is a key result of loop quantum cosmology, responsible for
an interesting phenomenology in the very early universe. We
demonstrate the way matter with arbitrary scale factor dependence 
in Hamiltonian
incorporates this change in its effective dynamics in the
loop modified phase. For generic matter, the  
equation of state starts varying near a critical scale factor, 
 becomes negative below it and violates strong energy condition.
This opens a new avenue to generalize various phenomenological applications 
in loop quantum cosmology. We show that different ways to define 
energy density may yield radically different results, especially for the case
corresponding to classical dust. We also discuss implications 
for frequency dispersion induced by modification to geometric density
at small scales.
\end{abstract}


\maketitle

\section{Introduction}

Quantum gravity is expected to radically modify our classical
intuition of space-time and matter. Progress in loop quantum
gravity (LQG), one of the background independent and non-perturbative
candidates for quantization of gravity, suggests that at the
quantum level classical spacetime continuum is replaced by
discrete quantum geometry \cite{lqg_review,geometrical_op}. The
continuum spacetime emerges from quantum geometry in a large
eigenvalue limit. An important question in this setting is the way
behavior of  the ordinary matter is modified at small scales.
Answering this question in complete generality is difficult since
we lack a full theory of non-perturbative quantum gravity
including matter. However, valuable insights can be obtained if we
work in a simpler symmetry reduced setting like quantum cosmology.

Loop quantum cosmology (LQC) is the  quantization of homogeneous and
isotropic mini-superspaces based on LQG whose
applications include a resolution of the big-bang
singularity  in homogeneous and
isotropic mini-superspace setting \cite{bigbang,singularity,Bohr} 
(for a recent review
see \cite{martin}, also see Refs. \cite{crit,crit1} for critical discussions). One of its key result is that eigenvalues of
geometrical density operator (or positive powers of inverse scale
factor in general) become proportional to positive powers of the scale
factor \cite{footnote2} below a critical value,
 $a_* = \sqrt{j \gamma/3} \lp$. Here $j$ is
a half integer greater than unity, $\lp$ is the Planck length and
$\gamma \approx 0.2375$ is the Barbero-Immirzi parameter whose
value is set by black hole thermodynamics \cite{bek_hawking1}.
Also as, $a \lra 0$ the spectrum of the inverse scale factor operator
remains bounded and curvature does not diverge.

At the fundamental level the evolution in LQC
is governed by quantum difference equations. However above the
scale factor $a_i \approx \sqrt{\gamma} \lp$ ,  dynamics can be
approximated  by Friedmann equations with non-perturbative
modifications \cite{Semi}. The spacetime does not immediately
become classical as we go from low to high eigenvalues of the scale factor
 and the regime $a_i \lesssim a \lesssim a_*$ is very
interesting to explore new physical effects. The dynamics in this
semi-classical regime has been studied extensively for a scalar
field and various interesting results have been obtained, like
dynamical initial conditions for the universe \cite{Initial},
naturalness of inflation \cite{superinflation,closedinflation,inflationcmb,
Robust,infbounce1,lidsey1,infbounce2,effham1}
, avoidance of big
crunch in closed models \cite{BounceClosed,BounceQualitative,effham2}, non-singular brane
bounce  in cyclic models \cite{Cyclic}, emergent universe scenario \cite{emergent}, possibility of
signatures in cosmic microwave background (CMB) \cite{inflationcmb} and discrete corrections to classical trajectories \cite{date2,eff_diff}. The non-perturbative modifications to Friedmann equations have also been shown to match well with underlying quantum
difference equations till very small scales \cite{eff_diff,time}. Investigations to the issues pertaining to perturbations have also 
been initiated \cite{golam1,golam2}. LQC techniques have also been applied to scalar field collapse models and it has been shown that
black hole and naked singularities can be avoided \cite{bhole,naked}.

Though LQC has yielded various interesting results in scalar field
dynamics, phenomenological applications for arbitrary matter
remain an open issue.
This problem is not only important by itself to gain insights in behavior of matter at
scales near and below $a_*$, but also to have understanding of a more complete model of the universe
with loop modifications. It should be noted that even in most of the
scalar field applications considered in LQC, various properties of the
scalar field are just assumed as in standard cosmology. Therefore, it
is highly desirable to look for alternatives to scalar field
phenomenology. We shall recall that in classical cosmology scalar
field is a very attractive entity since it can easily violate the
strong energy condition in the presence of a potential, which can lead to various interesting
consequences. Our interest here is to explore the possibilities for generic matter when scale factor
is smaller than $a_*$. If equation of state of matter like dust or
radiation behaves in a radically different way below some scale factor, then the results obtained 
in LQC using scalar field might be generalized to other matter. In particular any such result is particularly useful
to study the last stages of a contracting universe or gravitational collapse scenarios in LQC. Further profound insights
might be obtained for some phenomena in the very early universe, thus making the 
LQC phenomenology for generic matter very important.

Since  we are working in a symmetry reduced framework, 
incorporating generic matter configurations  is a difficult issue. This is due to the reason that
various forms of matter  in  classical cosmology, like perfect fluids, are not 
as fundamental as the scalar field. Further, consistent analysis of  perfect fluids 
at all scales (including discrete quantum regime) may require a full non-perturbative treatment of 
quantum gravity with matter. Since such a theory is still under development, our treatment here would
be very phenomenological. 

In classical cosmology matter coupling to gravity is identified via the stress-energy
tensor of a perfect fluid or the energy density and pressure. Latter are related via the thermodynamic relation of 
equation of state and obey
the conservation law derived from the adiabatic expansion of the universe (which is same as the divergence free property
of stress-energy tensor). The conservation law provides us with the proportionality of energy density with the scale factor.
Matter components with constant but different equation of states are proportional to different powers of the scale factor.
Since we can not include such classical fluids in current treatment, we would investigate 
the modifications to Hamiltonian with suitable scale factor dependence such that the corresponding energy density at classical volumes $(a \gg a_*)$ has the same behavior in scale factor as
the energy density of a classical matter component like dust, radiation (or relativistic gas of particles) etc. In this way
we can obtain  insights on the way energy density for matter shall modify as we approach semi-classical scales. In Ref. \cite{time}, it was
shown that inverse scale factor
modification to energy density of matter behaving as classical dust is sufficient to mimic the underlying 
quantum dynamics
to the scales of the order of $a_i$. Further at large volumes ($a \gg a_*$), we recover back
classical form of densities and standard dynamics. Thus, dynamical equations with modified energy density make a good
effective theory to the fundamental quantum difference equations in the semi-classical domain and we can study the
variation from classical behavior, like for the equation of state, in the effective description without 
referring to the underlying quantum dynamics. Our analysis would be
based on this effective phenomenological picture.

We emphasize that we do not include perfect fluids in current framework of LQC but we investigate matter Hamiltonians
whose energy density at classical scales mimic those of fluids like dust, radiation or stiff matter. We further assume as in
Ref. \cite{date2} that time scales of loop modified cosmological dynamics are large compared to those which establish
thermodynamic equilibrium in matter processes such that notion of equation of state is well defined and conservation law
can be used. It is possible that in deep quantum regime near $a_i$, above assumption may break down. In any case, in the
full quantum zone our phenomenological picture would be invalid and thus insights gained from this work can be
trusted only for scales not much below $a_*$ with the latter chosen large enough compared to $\lp$.

Our first result is to show that there are different ways to define energy density given a matter Hamiltonian, depending on whether we construct
a corresponding quantum operator or not. This is also related to the way we can obtain the effective
Friedmann equation in the semi-classical regime, either as an extension of the classical Friedmann equation or the 
semi-classical limit of the corresponding quantum construction.  
We demonstrate that the distinction between energy densities become important especially in the case for matter 
Hamiltonian independent of the scale factor. Phenomenologically this would correspond to
the coupling resembling classical dust. This can have important
consequences for the perturbations in the semi-classical regime. 
However, irrespective of the choice of definition of the energy density, we further show that
equation of state  shows variation from the behavior at classical volumes near and below $a_*$. This would effectively correspond to
existence of a negative pressure by using the conservation law. Thus matter which couples with classical gravity as pressureless
or positive pressure component, may transform into a negative pressure form at scales below $a_*$. Interestingly this is true even when the
equation of state is constant in classical theory, thus giving an indication that equation
of state may vary in semi-classical regime. As discussed earlier this
can have profound implications for scales $a \lesssim a_*$ with
the possibility that matter like dust and radiation can provide a
viable alternative to scalar field like in gravitational collapse scenarios \cite{bhole,naked}.
 This result is also important for multi-component models in semi-classical LQC. Our results are also immediately applicable to
models with more than one scalar field where one of the scalar field behaves effectively as dust or radiation with a suitable choice of potential.
Then the variation of equation of state to negative values for such
matter component suggests that may be it is possible to successfully generalize 
previous results of LQC. We also show that for
matter coupling corresponding to radiation, the frequency experiences
dispersion at scales smaller than $a_*$. Interestingly,  dispersion
is similar to the results obtained earlier using trans-Planckian cut offs \cite{planck1,planck2}.

\section{Modified Dynamics}

The root of modification of dynamics in LQC can
be traced back to the operator representing the quantum inverse scale
factor. Since the scale factor in LQC has
discrete eigenvalues including zero, one begins with an identity
on the Ashtekar-Barbero phase space \cite{thiemann_matter}. In
terms of the basic phase space variables, connection $c$ and the
triad $p$, $a^{-1}$ is given by \cite{ICGC} 
\be a^{-1} =
\bigg[\frac{3}{8 \pi G \gamma l} \, \{c,|p|^l \} \bigg]^{1/(2 - 2
l)} ~ \label{eq1}
\ee where $l$ is a quantum ambiguity parameter with $0 < l < 1$. The triad $p$
is related to the scale factor $a$ via $|p| = a^2$ and on
classical solutions (of general relativity) $c$ is given by $c =
(k - \gamma \dot a)/2$, $k$ being curvature index which
we take to be zero in this work. We can then quantize this
identity and obtain the eigenvalues of inverse scale factor. It
turns out that the eigenvalue spectrum is bounded on the entire
Hilbert space  and evolution through $a=0$ is non-singular
\cite{bigbang,singularity,Bohr}. The eigenvalues of ${\widehat{
(1/a)}}$ below $a_*$ become proportional to the positive powers of
scale factor and are not inverse of those of $\hat a$.

The geometrical density operator $(\widehat{1/a^3})$ can be
similarly constructed and its eigenvalues for large $j$ are
approximated as \cite{superinflation} 
\be
d_{j,l}(a) = D_l(q) \, a^{-3}, ~~ q := a^2/a_*^2, ~~ a_* := \sqrt{j \gamma/3} \, \lp \label{density}
\ee
where
\begin{eqnarray} \label{defD}
 D_l(q) &=&
\left\{\frac{3}{2l}q^{1-l}\left[\frac{1}{2 + l}
\left((q+1)^{l+2}-|q-1|^{l+2}\right)\right.\right. \\
&-& \left.\left.\frac{q}{1 + l} \left((q+1)^{l+1}-{\rm sgn}(q-1)
|q-1|^{l+1}\right)\right]\right\}^{\frac{3}{(2-2l)}} \nonumber.
\label{Deq}
\end{eqnarray}
Radical modifications to the behavior of geometrical density
become obvious if $a \ll a_*$ when
\be d_{j,l}(a) \approx
\left(\frac{3}{1+l}\right)^{3/(2-2l)}\,
\left(\frac{a}{a_*}\right)^{3(2-l)/(1-l)} a^{-3} ~.
\label{approxdensity} \ee
At classical scales $a \gg a_*$,
$d_{j,l} \approx a^{-3}$ and we recover back the classical
description.

For a matter specified by Hamiltonian ${\cal H}_M$, dynamics can
be obtained from total Hamiltonian constraint which on
quantization leads to the following difference equation
\cite{Bohr}
\bq
&&\hskip-.5cm \left(V_{\mu + 5 \mu_0} - V_{\mu + 3 \mu_0} \right)
\, \psi_{\mu + 4 \mu_0} - 2 \left(V_{\mu + \mu_0}  - V_{\mu -
\mu_0} \right) \,
\psi_{\mu} \\
&+& \left(V_{\mu - 3 \mu_0} - V_{\mu - 5 \mu_0} \right) \,
\psi_{\mu - 4 \mu_0} = - \frac{8 \pi G}{3} \, \gamma^3 \lp^2 \,
{\cal \hat H}_M(\mu) \psi_\mu \nonumber
\eq
where $V_\mu$ are eigenvalues of volume operator and are related
to eigenvalues ($\mu$) of triad as $V_\mu = (\gamma |\mu|/6)^{3/2}
\lp^3$. In the semi-classical limit it can be shown that above
equation can be approximated by the following differential equation
\cite{Semi,superinflation}
\be
- 3 \, \dot a^2 \, a + 8 \, \pi \, G {E}_M(a,\phi) = 0
~.\label{hamcons}
\ee
Here $E_M(a,\phi)$ are eigenvalues of matter Hamiltonian operator,
assumed to  correspond to a matter field $\phi$. 
It is important to note that if
${\cal H}_M$ depends on the inverse scale factor, then
$E_M(a,\phi)$ inherits the modifications to the eigenvalues of
$({\widehat {1/a}})$ upon quantization. Thus, dynamics in the 
semi-classical regime is distinct from its classical counterpart \cite{amb}.
A useful example is the case of
a massive scalar field whose classical Hamiltonian
\be
{\cal H}_M = \frac{1}{a^3} \, \frac{p_\phi^2}{2} + a^3 \, V(\phi)
\ee on loop quantization yields modified eigenvalues as
\cite{superinflation,closedinflation,inflationcmb}
\be E_M(a,\phi) = d_{j,l}(a)\,  \frac{p_\phi^2}{2} + a^3 \,
V(\phi) ~. \label{hamphi} \ee Dynamics in this case is then
completely determined by further using the modified Klein-Gordon
equation \cite{superinflation,closedinflation,inflationcmb} \be
\ddot \phi + \left(3 \frac{\dot a}{a} - \frac{\dot D_l(q)}{D_l(q)}
\right) \, \dot \phi + D_l(q) \, V_{,\phi}(\phi) ~ = 0 ~,
\label{kgeq} \ee where we have used the Hamilton's equations $\dot
\phi = d_{j,l}(a) p_\phi$ and $\dot p_\phi  = - a^3 \,
V_{,\phi}(\phi)$. In the regime $a \lesssim a_*$, $D_l(q) \ll 1$
and the modified dynamics becomes independent of the potential.
Also, the Klein-Gordon equation (\ref{kgeq}) can be approximated
as
\be
\ddot \phi  - 3 \bigg[\frac{2-l}{1-l} - 1 \bigg] \, \frac{\dot
a}{a} \, \dot \phi  ~ \approx 0 ~
\ee
where we have used eq.(\ref{approxdensity}). Since $0 < l < 1$ the
coefficient of $\dot \phi$  in eq.(\ref{kgeq}) changes sign
compared to its classical value and the scalar field experiences
anti-friction (friction) instead of friction (anti-friction) for
an expanding (contracting) universe. Such a peculiar change is
responsible for various interesting physical effects for example
super-inflation
\cite{superinflation,closedinflation,inflationcmb,Robust,infbounce1,lidsey1,infbounce2,effham1}, avoidance of big crunch
\cite{BounceClosed,BounceQualitative,effham2} and resolution of black hole and naked singularities with possible
observable signatures \cite{bhole,naked}.

Let us now determine the modifications to the energy density.
In classical theory once we know  $E_{M_{\mathrm{cl}}}(a,\phi)$, it
is straightforward to evaluate classical energy density which is
defined as $\rcl = E_{M_{\mathrm{cl}}}(a,\phi)/a^3$. Since
geometric density is modified in the semi-classical regime, the
notion of energy density becomes subtle. To obtain the
semi-classical density one way is to define a density operator
$\widehat{\rqm} = \widehat{{\cal H}_M/a^3}$ and then take the
semi-classical limit. Another way is to define
semi-classical density as the ratio of modified energy density to
the volume i.e. $\rsc = E_M(a,\phi)/a^3$ (where $E_M(a,\phi)$
includes the appropriate modifications to inverse scale factor)
\cite{superinflation}. Note that since we so far lack the
understanding of physical inner
product in LQC, these definitions are considered with the caveat that they may
not correspond to the expectation values taken with respect to
physical semi-classical  states. It may turn out that energy density
obtained via expectation values is different than
$\rqm$ or $\rsc$, which may require further analysis of results
presented here. There are other possible quantization ambiguities in defining energy density, like we 
can classically write $a^{-3} = a^{-3 \beta} a^{3(\beta - 1)}$ for $\beta > 0$, quantize it and obtain the 
energy density which would now depend on $\beta$. Note that all such quantization ambiguities lead to same classical energy density
and in a quantum theory a choice has to be made, either on the ground of natural value of parameter $\beta$ (which would be $\beta = 1$)
or the connection with full theory of LQG. The origin of such an ambiguity parameter is similar to the parameter $l$ which is present in 
eq.(\ref{eq1}) and is discussed in detail in Ref. \cite{Robust} where
it was shown that it is the natural choice of parameter which may
arise from LQG. This augurs well with the general expectation that natural values of such quantization ambiguity parameters would be favored
by the full theory of LQG. We would focus our discussion on the natural
value of the $\beta$ parameter and discuss the implications for other possible values in the concluding section. It would turn out that
physics and conclusions of this work are very robust to arbitrary choices of $\beta$.

The eigenvalues for $\widehat{\rqm}$  can be
obtained in the way  described above for
$\widehat{\cal H}_M$ and $\widehat{1/a^3}$ and they are
\be
\rqm = d_{j,l}(a) \, E_M(a,\phi) = D_l(q) \, a^{-3} \, E_M(a,\phi)
= D_l(q) \, \rsc ~. \label{denrel}
\ee
For $a \lesssim a_*$, $D_l(q) \lesssim 1$ implies $\rqm \lesssim
\rsc$. It is to be noted that $\rqm$ incorporates modifications
both in energy and geometric density eigenvalues, whereas $\rsc$
does not receive any contribution from modifications in behavior
of $1/a^3$.
Though dynamics does not depend on the choice of 
the density, there are important distinguishing features which we discuss in next section.

Modifications to energy and density eigenvalues of matter Hamiltonians with
arbitrary scale factor dependence can be determined in a similar way. In the semi-classical regime, at the effective level, we shall first replace the inverse
scale factors in Hamiltonian with the appropriate powers of $d_{j,l}$ and then 
obtain energy density $\rqm$ (or $\rsc$). For example, if ${\cal H}_M$ is independent of the scale factor 
then its energy eigenvalues would not get modified for scales less than $a_*$. The energy density defined via $\rqm$ is modified. At classical scales
energy density becomes proportional to $a^{-3}$. Thus, this form of matter coupling would resemble that of 
dust at large volumes. Similarly, the Hamiltonian whose energy density would classically couple as that of radiation
is ${\cal H}_M(a) \propto 1/a$. The energy eigenvalues in this case get modified below $a_*$, which is equivalent to
modification to behavior of frequency at small volumes (see Ref. \cite{golam2} for a similar discussion).

The semi-classical dynamics can be obtained directly from the Hamiltonian constraint eq.(\ref{hamcons}) and 
since the latter does not depend on $\rqm$ or $\rsc$, the trajectories like $a(t)$ would be identical for both choices
of energy density. However, if instead of using Hamiltonian constraint we wish to use effective Friedmann equation to
determine dynamics then there are some subtleties. One way  to obtain the effective Friedmann equation is as done 
classically, that is dividing the Hamiltonian constraint by $a^3$. This yields the gravitational energy density as 
$\dot a^2/a^2$ which is identified with the square of the classical Hubble rate $(H_{\mathrm{cl}} := \dot a/a)$ 
and is equal to the classical matter energy density 
(up to numerical factor of $8 \pi G/3$). With this algorithm, the division of eq.(\ref{hamcons}) 
by $a^3$ yields the modified Friedmann equation
with unmodified gravitational energy density, and thus effective
Hubble rate which is same as the classical one,
i.e. $H_{\mathrm{sc}}^2 = H_{\mathrm{cl}}^2 := \dot a^2/a^2$, and is 
proportional to $\rsc$. 

An alternative method to obtain semi-classical effective Friedmann equation is to first quantize the classical
Friedmann equation and then obtain its semi-classical limit \cite{hubble}.
As explained  for the construction of $\widehat{\rqm}$, this method would imply that the modified 
Friedmann equation in the regime $a \lesssim a_*$ is given by $d_{j,l}$ multiplied with eq.(6). This means that 
the gravitational energy density also gets modified below $a_*$ and is given by $D_l \, \dot a^2/a^2$. That is the effective 
Hubble rate in semi-classical regime is given by $H_{\mathrm{q}} := D_l^{1/2} \, H_{\mathrm{sc}} = D_l^{1/2} \, \dot a/a$, 
whose square is proportional to the energy density $\rqm$. For $a > a_*$, both $H_{\mathrm{q}}$ and $\rqm$ approach 
$H_{\mathrm{cl}}$ and $\rcl$ respectively, and the classical dynamics is recovered. Interestingly the modified Friedmann
equation for $H_{\mathrm{q}}$ can be further divided by $D_l$ to obtain a $\dot a^2/a^2$ equation which resembles the
modified Friedmann equation obtained from the first method, however that does not imply that either $\dot a^2/a^2$ 
or $\rsc$ can be identified with the actual gravitational and matter energy densities ($D_l \, \dot a^2/a^2$ and $\rqm$). 

The point to note is that different gravitational and matter densities appear in modified Friedmann equation depending
on the way latter is obtained, either by extending classical theory to semi-classical regime which leads to an unmodified Hubble rate 
or taking the semi-classical limit of the quantum theory which modifies classical Hubble rate to $H_{\mathrm{q}}$.
As we mentioned above, these differences (or ambiguities to obtain effective Friedmann equation) 
do not affect dynamical trajectories and the phenomenological applications like inflation and bounces in LQC are robust
to such ambiguities. However they do lead to 
different effective Hubble rates and energy densities in the semi-classical regime $a \lesssim a_*$.
This can be important for example in the investigations of constraints on the value of loop parameters 
to yield viable initial conditions for conventional inflation \cite{Robust}. We recall that in Ref. \cite{Robust}, $H_{\mathrm{sc}}^{-1} > \sqrt{\gamma} \lp$ was used
to obtain constraint 
on parameter $j$. The 
phenomenological constraint on $j$ is thus expected to change if instead we use  $H_{\mathrm{q}}$.
Further, in Ref. \cite{eff_diff} estimates on the scale below
which discrete quantum geometry corrections become significant to the dynamics have been obtained. This scale which is
related to a critical density obtained using $\rsc$ may also be affected if we use $\rqm$. These issues will
 important for future investigations in this direction.

\section{Variation of equation of state}

Given classical matter  with a constant equation of
state $(\wcl)$, matter density evolves as
\be
\rcl = \rho_0 \, a^{-3 (1 + w_{\mathrm{ cl}})} \label{classden}
\ee
where $\rho_0$ is a constant. At the effective phenomenological level the 
modifications to energy density of matter as for $\rqm$ corresponds to 
replacing inverse powers of $a$ in
eq.(\ref{classden}) with appropriate powers of $d_{j,l}$. Thus
\be
\rqm = \rho_0 \, d_{j,l}^{~(1 + \wcl)} = D_l^{(1 + \wcl)} \,
\rcl~. \label{den_mod}
\ee
Further, on using eq.(\ref{denrel}) we obtain
\be
\rsc = D_l^{-1} \, \rqm  = D_l^{\wcl} \, \rcl~. \label{den_modsc}
\ee
Hence both prescriptions ($\rqm$ and $\rsc$) to obtain
semi-classical density lead to the modifications from classical
density. The energy conservation law obeyed by matter immediately ensures
that pressure must change to accommodate any modification in energy density. The rate of change of
energy density $\rqm$ with respect to the scale factor is 
\be
a ~ \frac{d}{d a} \rqm = (1 + \wcl) \, \rqm \bigg[\frac{d \ln D_l}{d \ln a}  - 3
 \bigg] ~.
\ee
Using this in energy conservation equation
\be
a \, \frac{d}{d a} \rho = - 3 \, (\rho + p) ~ \label{energycons}
\ee
we easily obtain the
expression for modified pressure $p_{\mathrm{q}}$,
\be
p_{\mathrm{q}} = \wcl \, \rqm - (1 + \wcl) \, \frac{d \ln D_l}{d \ln a} \frac{\rqm}{3}
\ee
which leads to an effective equation of state ($\wqm$) defined as
the ratio $p_{\mathrm{q}}/\rqm$,
\be
\wqm = \bigg[\wcl - \frac{(1 + \wcl)}{3} \, \frac{d \ln D_l}{d \ln a} \bigg] ~. \label{wqm}
\ee
Similar derivation can be done by starting from
energy density $\rsc$ and the resulting effective equation of state $\wsc =
p_{\mathrm{sc}}/\rsc$  is
\be
\wsc = \wcl \, \bigg[1 - \frac{1}{3}\, \frac{d \ln D_l}{d \ln a} \bigg] ~. \label{wsc}
\ee

At classical scales $a \gg a_*$, $D_l(q) = 1$. However at scales 
near and below $a_*$, $D_l(q)$ starts varying and becomes much smaller
than unity. This leads to the variation of equation of state for matter 
for $a \lesssim a_*$. If our phenomenological picture is allowed to be trusted
even for scales $a \sim a_i$, then 
fixing the parameter $l$ to its natural value of $3/4$
\cite{Robust}, it is straightforward to see from eq.(\ref{Deq})
that for $a \sim a_i \approx \sqrt{\gamma} \lp$, $d \ln D_{3/4}/d \ln a \approx 15 $ and thus 
\be
\wqm \approx -4 \wcl - 5
\ee
and
\be
\wsc \approx -4 \wcl ~.
\ee
As discussed earlier we do not expect this behavior to be 
valid at scales so close to Planck length, however we do expect that 
for scales close to $a_*$, equation of state would start varying
even if it is constant for $a \gg a_*$. We illustrate this with
examples of matter Hamiltonians of the form $C_1/a^3$, $C_2/a$ and $C_3$
where $C_i$ are some constants. Since the energy density of these Hamiltonians
resemble that of stiff matter, radiation and dust respectively in scale factor dependence at classical volumes,
we would refer to examples with this correspondence. We would also review the case of massive scalar field
and study the effect of using $\rqm$ instead of $\rsc$ as done earlier \cite{superinflation,inflationcmb,Robust,infbounce1,
infbounce2}.

\begin{figure}
\begin{center}
\includegraphics[width=9cm,height=7.5cm]{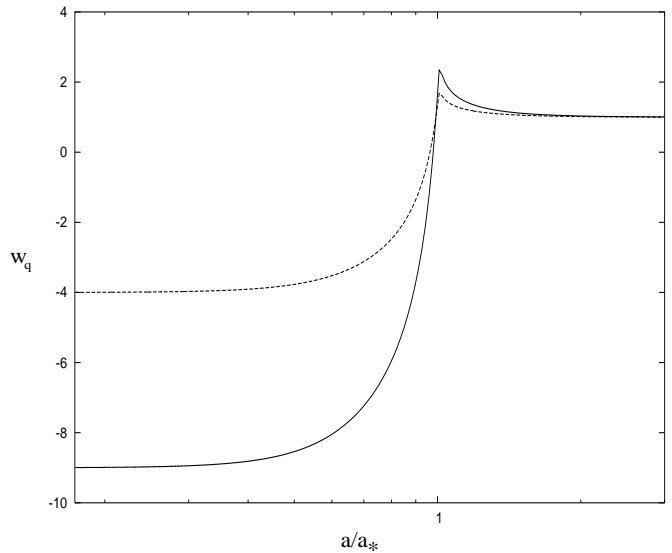}
\end{center}
\vskip-0.5cm \caption{Behavior of effective equation of state for
matter whose energy density at classical volumes has the behavior of that of stiff matter. The solid curve
shows $\wqm$ and dashed curve shows $\wsc$. Parameters are $j =
100$ and $l = 3/4$.  Same variation holds
for a massive scalar field (for details see text).} \label{fig1}
\end{figure}

{\underline {\it Stiff Matter:}} For matter Hamiltonian which is of the form
$C_1/a^3$, both $\rqm$ and $\rsc$ approach $\rcl \propto a^{-6}$ as for stiff matter at scales $a \gg a_*$.
As expected the equation of state $\wqm$ or $\wsc$ are equal to unity at large volumes. However, for scales
$a \sim a_*$, it starts varying and becomes negative for $a < a_*$. The increase in equation of state
from its classical value for a small domain near $a_*$ is due to the corresponding peak in
$D_l(q)$ for $a \sim a_*$. The variation of effective 
equation of state is shown in Fig. \ref{fig1}. As can be seen, the equation of state quickly becomes less than $-1/3$
after the peak at $a \sim a_*$, that is it violates strong energy condition.  It also becomes less than $-1$ for values of scale factor which are of the order of $0.8 a_*$, thus violating weak energy condition. 
This behavior is independent of choice of loop quantum parameters and suggests that matter
with a constant stiff equation of state at classical volumes may transform effectively into a form with
an equation of state which is negative. Though of little validity for our phenomenological picture we see that 
for $a \sim a_i$, $\wqm \approx -9$ and $\wsc \approx -4$.

\begin{figure}
\begin{center}
\includegraphics[width=9cm,height=7.5cm]{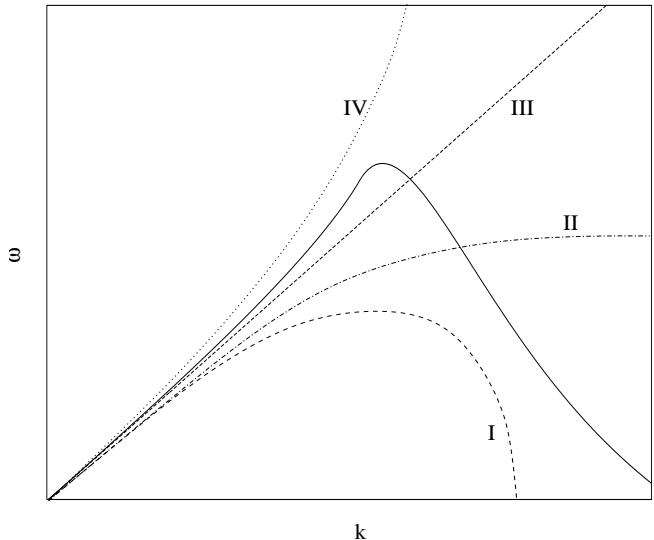}
\end{center}
\vskip-0.5cm \caption{Evolution of frequency $\omega$ (solid curve) with $k$. We choose quantum gravity parameters as $l=3/4$ and $j=100$. Behavior for other choices is similar. Classical curve denoted by III is not modified at any value of scale factor. The curves I,IV and II are obtained respectively in Ref.\cite{planck2} and Ref.\cite{planck1} by  cutoffs to dispersion relation at short scales. Quantum gravity naturally produces a modification to frequency at short scales without introduction of any cutoff.} \label{fig2}
\end{figure}

{\underline {\it Radiation:}} Matter Hamiltonians which are proportional to 
inverse scale factor lead to energy density of the form of radiation at classical
volumes. The Hamiltonian gets modified by $d_{j,l}^{1/3}$
in the semi-classical regime. Effectively, it implies that
frequency corresponding to classical radiation component gets modified
from the standard behavior below $a_*$. 
In this case classically we have $E_{M_{\mathrm{cl}}} = E_0 a_0/a$ (with $E_0$ and $a_0$ being constants) which gets modified to
$E_M = d_{j,l}^{1/3} a_0 E_0 = D_l^{1/3} a^{-1} a_0 E_0$. Thus for $a \lesssim a_*$, $E_M$ would become proportional to positive powers
of scale factor. Since for radiation $E_M$ is linearly related to frequency via Planck law, this modification implies
a change in behavior of frequency at scales below $a_*$ which would be given by 
$\omega = \omega_0 a_0 d_{j,l}^{1/3} = \omega_0 a_0 D_l^{1/3} a^{-1}$. Modifications to the behavior of frequency at 
short scales have been 
expected and desired from quantum gravity models. However, most of the times these are introduced by introduction of a short scale cut off \cite{planck1,planck2}. Here we see that such a cut off is provided by the scale below which
behavior of density changes. We have plotted the behavior of $\omega$
with $k = 2\pi/\lambda$ in Fig. \ref{fig2}. As can be
seen, the modification to frequency is similar to those expected earlier and inspires further investigations to understand
modification to dispersion relation using quantum gravity models as we
have done here. These issues will be addressed elsewhere.

\begin{figure}
\begin{center}
\includegraphics[width=9cm,height=7.5cm]{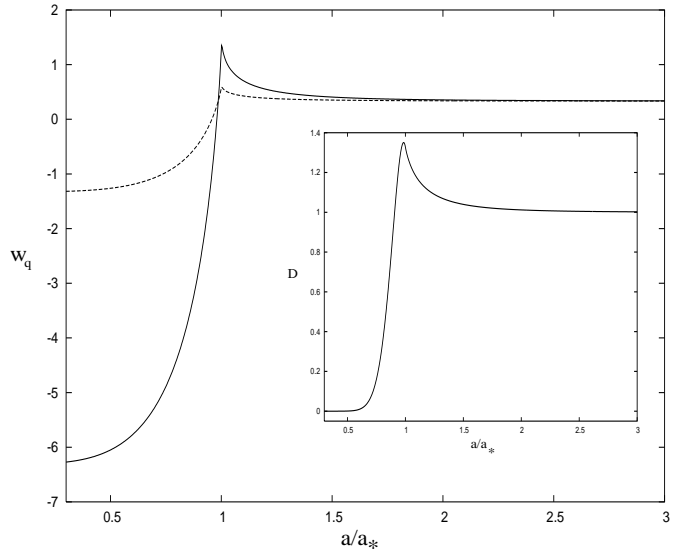}
\end{center}
\vskip-0.5cm \caption{Evolution of $\wqm$ (solid curve) and $\wsc$
(dashed curve) for ${\cal H}_M \propto a^{-1}$ with loop
parameters as in Fig. \ref{fig1}. If phenomenological picture
is valid till $a \sim a_i$, matter which couples to gravity at classical volumes as radiation 
transforms into one with super-negative pressure. 
Inset shows the variation of $D_l(q)$ with its peak at
$a \sim a_*$.} \label{fig3}
\end{figure}

The equation of state for matter coupling like radiation via $\rqm$ construction becomes
\be
\wqm = \frac{1}{3} - \frac{4}{9} \, \frac{d \ln D_l}{d \ln a} ~.
\ee
At classical scales, $D_l(q) = 1$  and thus $\wqm = \wsc = 1/3$. However
for scales less than $a_*$, $D_l(q)$ starts varying which leads to variation 
of $\wqm$ and $\wsc$. In Fig.
\ref{fig3}, we have shown the evolution of $\wqm$ and $\wsc$ with
scale factor. Inset shows evolution of $D_l(q)$ with its peak at
$a \sim a_*$. As in the the case of stiff matter, the variation in
$\wqm$ is more rapid than $\wsc$. Though both variations suggest that
strong and weak energy conditions are violated for scales near $a_*$, 
these violations occur at slightly larger scale factors for $\wqm$ than 
for $\wsc$.

\begin{figure}
\begin{center}
\includegraphics[width=9cm,height=7.5cm]{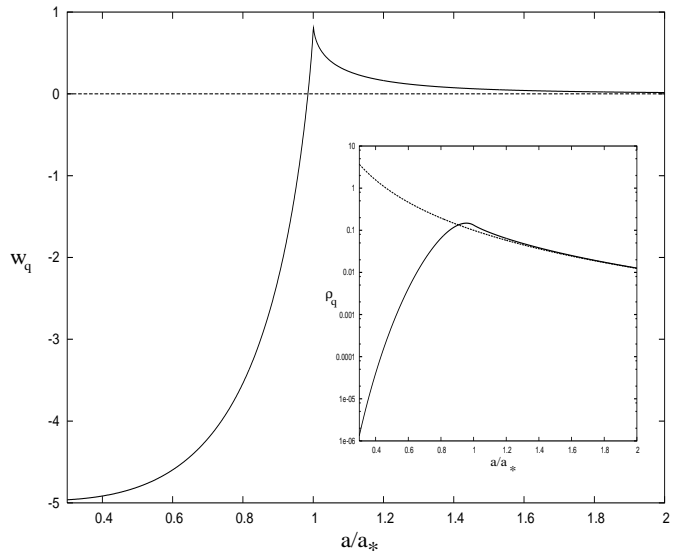}
\end{center}
\vskip-0.5cm \caption{Variation of effective equation of state for
matter satisfying $\wcl = 0$ in the classical regime.
Solid curve depicts $\wqm$ and dashed curve shows $\wsc$.
Parameters are same as in Fig. \ref{fig1}. Inset shows evolution
of $\rqm$ with scale factor which is well regulated as $a \lra 0$
whereas the energy density without $d_{j,l}$ modification blows up
(shown by dashed curve).} \label{fig4}
\end{figure}

{\underline {\it Dust:}} Hamiltonian which is independent of the scale factor
does not receive any modifications due to $d_{j,l}$ for scales less than
$a_*$. The energy density $\rsc$ is equal to $\rcl$ at all scales and thus diverges
for small scale factors. However, on using density operator we obtain
 $\rqm = D_l \rcl$ and
\be
\wqm =  - \frac{1}{3} \, \frac{d \ln D_l}{d \ln a}.
\ee
We have shown the variation of effective equation of state in Fig.
\ref{fig4}. As for the case of stiff matter and radiation, $\wqm$
varies for $a \sim a_*$ and becomes negative below the critical scale factor.
It also violates weak energy condition for $a \sim 0.9 a_*$. On the other hand $\wsc$ remains constant
(equal to zero) all through the period of evolution.
The differences between $\rqm$ and $\rsc$  become very
evident in this case. 
As $a \lra 0$, $\rsc$
blows up whereas $\rqm$ remains finite. It is important to note that in $\rqm$
effects of geometric density regulate the diverging energy density
at small scale factors.

{\underline {\it Massive Scalar Field:}} The case of massive scalar field has been studied in detail
in LQC \cite{superinflation,closedinflation,inflationcmb,Robust,infbounce1,lidsey1,infbounce2,effham1,
BounceClosed,BounceQualitative,effham2,Cyclic,date2,golam1,golam2,emergent}. The period of super-acceleration in the regime $a \lesssim a_*$ 
plays a dominant role in most of the interesting effects like setting up right conditions for inflation or preventing
singularities. This phenomena also indicate that effective equation of state becomes less than -1 and thus weak energy condition is violated. 
The energy density of a
massive scalar field $\phi$ with potential $V(\phi)$ via density
operator construction is given by
\be
\rqm = d_{j,l} \, E_M(a,\phi) = \frac{\dot \phi^2}{2} + D_l(q) \, V(\phi) ~ \label{phiden}
\ee
where we have used eqs.(\ref{denrel}) and (\ref{hamphi}).
The expression for effective pressure can be obtained by using eq.(\ref{energycons}) and Klein-Gordon equation (eq.(\ref{kgeq})). It
turns out to be
\be
p_{\mathrm{q}} = \bigg[1 - \frac{2}{3} \, \frac{d \ln D_l}{d \ln a}
\bigg] \, \frac{\dot \phi^2}{2} - D_l(q)\, V(\phi) - \frac{1}{3} \, \frac{d \ln D_l}{d \ln a}\, V(\phi) ~
\ee
and thus effective equation of state can be obtained by using $\wqm = p_{\mathrm
q}/\rqm$. It shall be noted that effective negative pressure obtained from $\rqm$ is much stronger than that obtained from $\rsc$
leading to 
\cite{superinflation,infbounce1}
\be
\wsc = - 1 + \frac{2 \dot \phi^2}{\dot \phi^2 + 2 D_l(q) V(\phi)} \, \bigg[1 - \frac{1}{6} \,  \frac{d \ln D_l}{d \ln a}\bigg] ~.
\ee
Since $D_l(q) \ll 1$ for $a \ll a_*$, the potential terms become
negligible compared to kinetic terms and it is easy to verify that
both $\wqm$ and $\wsc$ behave in the same way as for stiff
matter. The variation of equation of state and the super-negative
pressure are thus as shown  in Fig. \ref{fig1}. It is important to note that use of $\rqm$ leads to stronger violation of
energy conditions than $\rsc$. As we discussed in the previous section
this issue might be  linked to the problem of stability of perturbations in the regime $a \lesssim a_*$ \cite{golam2}. Hence
choice of $\rqm$ or $\rsc$ is bound to play an important role in obtaining viable density perturbations in loop quantum modified regime.
We would leave such an investigation for future work.

\section{Conclusions}

Working in the semi-classical limit of LQC we have studied the
behavior of matter Hamiltonians  with arbitrary scale factor dependence, in particular those whose energy density scales as that
of dust, radiation and stiff matter at classical volumes. Through our phenomenological effective treatment we are able to gain useful
insights on what shall be expected to be the modified behavior of matter at scales near and less than $a_*$. Our first result is to show that
there are at least two ways to define energy density and thus equation of state. Both of them lead to same classical behavior for $a \gg a_*$, but for
$a \lesssim a_*$, there are significant distinctions between them. For
example in the case of matter Hamiltonian with no scale factor
dependence (classically corresponding to dust) , at small scale
factors energy density defined via $\rsc$ blows up whereas $\rqm$
remains regulated and finite. 

The effective equation of state for matter mimics the classical
behavior for $a \gg a_*$. However we have shown that near $a_*$ it starts varying even if
it is classically constant. It increases initially for $a \sim a_*$ and then rapidly decreases leading to violation of energy conditions.
This violation is independent of the choice of $\wqm$ or $\wsc$. It suggests that classical matter may effectively 
metamorphose itself to various forms at short scales and may serve as
viable alternative to scalar field phenomenology.
The case of radiation offers a new insight on the trans-Planckian
modifications to the frequency dispersion. It is intriguing that
inverse scale factor modifications may provide a natural explanation
to much sought frequency dispersion at short scales, however this
requires a detailed analysis which is beyond the scope of present discussion.

As we discussed in Sec. II, there is a new quantization ambiguity which may arise in construction of energy density via quantum
operator. It can be checked that if instead of taking $\beta = 1$, we keep it arbitrary positive value then $\rqm$ becomes
$\rqm = D_l^{\beta(1 + w_{\mathrm{cl}})} \, \rcl$  and the factor $(1 + \wcl)$ in eq.(\ref{wqm}) gets multiplied by $\beta$.  Since $\beta$ is positive, its multiplication with $(1 + \wcl)$ in eq.(\ref{wqm}) does not affect the qualitative behavior of equation of state $\wqm$.
For any choice of $\beta$, the effective metamorphosis of equation of state at scales smaller than $a_*$ would occur. 
The parameter $\beta$
only affects the magnitude of variation of $\wqm$. For example, if we
fix $l = 3/4$ then for the case of dust all values of $\beta > 1/15$ would imply violation of
strong energy condition for $a \sim a_i$.  By taking the same value of
$l$ and $\beta >1/5$, $\rqm$ for dust like matter would scale as 
positive power of scale factor for $a \sim a_i$ and energy density does not blow up at small scale factors. 
Similarly for radiation and
stiff matter, an arbitrarily chosen $\beta$  results in a different  magnitude of variation of the energy density and 
equation of state and the qualitative picture does not change. The result of modification to
frequency dispersion for radiation is independent of the choice of energy density and is unaffected by this ambiguity parameter.
This leads us to the conclusion that phenomenological effects discussed in this work are very robust and the qualitative picture 
does not depend on different choices of the quantization ambiguity parameter $\beta$. We recall that in Ref. \cite{Robust}, it was
demonstrated that phenomenological results are qualitatively independent of the choice of parameter $l$. We have earlier discussed that
parameter $\beta$ originates in a very similar way as $l$. Now we  see that both parameters also have a very similar 
effect towards phenomenological description and the underlying physical predictions are quite robust to the choices of these
parameters.

Future investigations
with full LQG techniques would be able to confirm or rule out
the expectations of metamorphosis of equation of state and natural modifications to frequency dispersion.
 This opens a novel avenue to explore phenomenology at $a \lesssim a_*$, 
with matter like dust and radiation.
Since the equation of state for matter coupling as classical dust or radiation
to gravity becomes negative for $a < a_*$, it implies that multicomponent models of scalar field interacting with various matter components
would also yield similar qualitative results like super-acceleration and bounce in semi-classical LQC. Of course
the results obtained here are directly applicable to loop quantum
cosmological models with two or more scalar fields where at least one
of them behaves as classical matter component like dust or radiation
with appropriate choice of potential. In Ref. \cite{bhole} and
\cite{naked} gravitational collapse scenarios with scalar field were studied and possibilities of interesting astrophysical signatures 
have been reported. However, in a more realistic scenario inclusion of matter which behaves as dust, radiation or stiff matter is
important to investigate the last stages of astrophysical collapse. Our results would be particularly useful in this arena and 
may open the possibility to link LQC phenomenology with astrophysical observations.

Though the choice of energy density does not affect the dynamical trajectories, it may however change some of the
phenomenological constraints imposed on loop parameters using Hubble rate (or energy density). 
This issue should be investigated further which may give us useful insights.
However, a more fundamental understanding of matter in LQC and detailed analysis
of physical semi-classical states might also guide us towards resolving this ambiguity.
Our result of the effective state metamorphosis may also have some interesting implications for the
problem of dark energy where some of the  scenarios require effective equation of state
to become less than $-1$. In particular recently considerable attention has focused on crossing the $-1$ divide in equation of state
(which separates domains of validity of weak energy condition) and the role of quantum gravity effects \cite{wec}. 
In semi-classical LQC such a behavior is observed very naturally for various forms of matter. Investigation 
on the relation of LQC dynamics with that of standard cosmology using a scale factor duality has been done earlier \cite{lidsey1}. 
On a speculative side such a duality might provide a valuable link between  quantum gravity effects at short scales with dark energy at large scale
factors of the universe.

{\bf Acknowledgments:} We are grateful to A. Ashtekar for a
careful reading of the manuscript,  extensive comments and helpful
discussions. We also thank M. Bojowald, G. Date, R. Maartens, 
E. Papantonopoulos, M. Sami, A. Toporensky, S. Tsujikawa, K. Vandersloot
and  G.V. Vereshchagin for discussions and comments.
Author's work is supported in part by Eberly research funds of
Penn State and by NSF grant PHY-00-90091.


\begin{thebibliography}{10}
\bibitem{footnote2} The notion of scale factor is slightly different than as used in
terms of conventional cosmology. In LQC it is related to the
cubic root of the volume of the fiducial cell necessary to define symplectic structure
for quantization and  is invariant to conventional rescaling freedom \cite{Bohr}.
\bibitem{amb} There are various possible quantization ambiguities in obtaining the semi-classical description, however their
effect on phenomenology is rather weak. For their detailed
discussion and robustness of phenomenology  we refer the reader to Ref. \cite{Robust}.
\bibitem{hubble} A construction of the Hubble operator on similar
  lines has been proposed earlier \cite{hubble1}, however subtle
  issues pertaining to dynamics and energy density were not addressed.
\end{thebibliography}
\end{document}